\documentclass{elsart}
\usepackage{graphicx,amssymb}
\journal{Journal of Magnetism and Magnetic Materials}
\begin{document}
\input{epsf}
\begin{frontmatter}
\title{An ab-initio theoretical investigation of the soft-magnetic properties
of permalloys}
\author[warw]{S.Ostanin\thanksref{now}}
\thanks[now]{Present address: Department of Earth Sciences, University College London,
Gower Street, London WC1E 6BT}
\author[warw]{J.B.Staunton\corauthref{cor}}
\corauth[cor]{Corresponding author.}
\ead{j.b.staunton@warwick.ac.uk}
\author[kuw]{S.S.A.Razee}
\author[mes]{B.Ginatempo}
\author[mes]{Ezio Bruno}
\address[warw]{Department of Physics, University of Warwick, Coventry CV4 7AL, U.K.}
\address[kuw]{Department of Physics, University of Kuwait, SAFAT 13060, Kuwait}
\address[mes]{Dipartimento di Fisica and Unit\`a INFM, Universit\`a di Messina,
         Salita Sperone 31, I-98166 Messina, Italy}

\begin{abstract}
 We study Ni$_{80}$Fe$_{20}$-based permalloys
 with the relativistic spin-polarized
 Korringa-Kohn-Rostoker electronic structure method.
 Treating the compositional disorder with the coherent potential 
 approximation, we investigate how the magnetocrystalline anisotropy, $K$, 
 and magnetostriction, $\lambda$, of Ni-rich Ni-Fe alloys vary with the addition of
 small amounts of non-magnetic transition metals, Cu and Mo. From our 
 calculations we follow the trends in $K$ and $\lambda$ and find the 
 compositions of Ni-Fe-Cu and Ni-Fe-Mo where both are near zero. 
 These high permeability compositions of Ni-Fe-Cu and Ni-Fe-Mo 
 match well with those discovered experimentally. We monitor the connection
 of the magnetic anisotropy with the number of minority spin electrons 
 $N_{\downarrow}$.  By raising $N_{\downarrow}$ via artificially increasing 
the band-filling of Ni$_{80}$Fe$_{20}$, 
 we are able to reproduce the key features that underpin the magnetic 
 softening we find in the ternary alloys. The effect
 of band-filling on the dependence of magnetocrystalline anisotropy on
 atomic short-range order in Ni$_{80}$Fe$_{20}$ is also studied. Our 
 calculations, based on a static concentration wave theory, indicate that the 
 susceptibility of the high permeability of the Ni-Fe-Cu and Ni-Fe-Mo alloys 
 to their annealing conditions is also strongly dependent on the alloys' 
 compositions. An ideal soft magnet appears from these calculations.
\end{abstract}
\end{frontmatter}

\section{Introduction}
     For a magnet to be soft, weak fields must be able to change the 
 overall magnetisation readily. Such high permeability comes about because
 the energies of the domain walls are low so that the magnetic
 anisotropy constants, $K$, are also very small.
 The internal stresses caused by the changes of magnetisation
 must also be minimal making the magnetostriction coefficients,
 $\lambda$, as small as possible.\cite{Tremolet}
 Roughly the magnetic permeability is proportional to $M_{s}^2/ K_{eff}$, 
 where $M_s$ is the saturation magnetisation and $K_{eff}$ is a 
 measure of magnetic anisotropy and magnetostriction, 
 $K_{eff} = K_{eff} (K, \lambda)$.

 Some of the magnetically softest ferromagnetic materials are based on the
 Ni$_{80}$Fe$_{20}$ permalloy and their high permeability finds them
 numerous 
 applications in, for example, magnetic recording heads, spin 
 valve devices and electrical power generation. 
 In binary Ni$_{1-c}$Fe$_{c}$ alloys, $K$ and $\lambda$ both vary with 
 concentration $c$ and in the vicinity of $c=0.2$ pass through zero 
 at slightly different compositions and so the addition of a third or fourth 
 component is required to achieve 'the focus of zero'.~\cite{Chikazumi}
 General guidelines for achieving near zero anisotropy and magnetostriction 
 in ternary and quaternary Ni$_{80}$Fe$_{20}$-based permalloys are now
 well established thanks to experimental work on hundreds of 
 samples.~\cite{Wakiyama,Pfeifer,Major} 

 Recently we have carried out a theoretical investigation for a
 prototypical soft magnetic metal, body-centered cubic (bcc) 
 iron~\cite{OSRD-PRB-2004}, and have examined how its $K$ and $\lambda$ would 
 vary if its lattice spacing (volume) and number of valence electrons (band-filling) 
 could be altered. We have found that on reducing 
 the band-filling and increasing the volume, iron's magnetic 
 properties soften considerably. This situation can be 
 realised in practice by doping bcc Fe with vanadium. We then tested our
 model by an explicit study of iron-rich Fe-V alloys and found the
 optimal composition for the smallest $K$ and $\lambda$. 
 In particular, we found that Fe$_{0.9}$V$_{0.1}$ is a high 
 permeability material. Good agreement with experimental 
 values for the magnetocrystalline anisotropy energy (MAE) and 
 magnetostriction of both Fe and \underline{Fe}V was found.

 In this paper, we start instead from a binary component `parent' soft magnet,
 Ni$_{1-c}$Fe$_{c}$ with $c \approx 0.2$ and see if a similar, ab-initio
 approach can explain how the addition of small amounts of non-magnetic 
impurities, Cu and Mo, produce the magnetically softest permalloys. 
 Our approach is based on the 
 spin-polarized relativistic Korringa-Kohn-Rostoker (SPR-KKR) density 
 functional theoretical (DFT)
 method, within the coherent-potential approximation (CPA)
 \cite{Razee+97,Razee+99} for randomly disordered alloys.
 We explore the variation of the two main factors, $K$ and $\lambda$,
 in Ni$_{1-c}$Fe$_{c-x}$Cu(Mo)$_{x}$ with concentrations $c$ and $x$
 and find the values which bring both quantities to optimally low values
 to promote high permeability. James {\it et al.}~\cite{James} have also
 recently carried out studies of the magnetostriction and magnetic anisotropy
 of 3d transition metal alloys from a `first principles' electronic 
 structure framework. From their calculations, they propose that the trends
 are linked to the number of occupied minority spin d-states. We see if
 such a link is evident from our calculations here on binary and ternary
 permalloys. 

 The magnetic anisotropy constants $K$ of Ni-rich permalloys 
 depend also on the existence of the long- and  short-range order. 
 There is a $L1_2$(Cu$_3$Au)-ordered Ni$_3$Fe phase below 600$^{\circ}$C. 
 The high permeability values of technically important 80~\% Ni permalloys 
 are achieved by specified annealing in the ordered range Ni$_3$Fe.
 Here, the manufacturing processes and final heat treatment of permalloys 
 are very important role for the alloys' microstructures and also their final
 magnetic properties (coercivity, hysteresis loop shape etc.). Consequently,
 we also present here a study of the effect of compositional order on the 
 high permeability Ni$_{80}$Fe$_{20}$ alloy and how this may be expected to
 vary when a non-magnetic transition metal dopant is added. This builds on our
 earlier work in which we described why directional chemical order and 
 significant uniaxial magnetic anisotropy is induced in permalloy when it is
 annealed in a magnetic field.~\cite{Razee+99} 
 
 In the next section we briefly summarise our approach for calculating 
 the MAE of metals and alloys, using the SPR-KKR method. 
 We then show how the magnetostriction can also be found. 
 In the following section we use the SPR-KKR method to study Ni$_{1-c}$Fe$_{c}$,
 ($c \approx 0.2$) and related Ni-Fe-Cu and Ni-Fe-Mo ternary alloys focussing on
 the region where high permeability appears. We discuss these results in terms 
 of band-filling (number of valence electrons per atom) and the number of 
 minority spin electrons plus features from the electronic densities of states. 
 We interpret a key feature of our results by a further investigation of 
 Ni$_{80}$Fe$_{20}$ in which the effect of doping with a non-magnetic impurity
 is modelled simply. The next section
 outlines the theoretical framework for compositional order in alloys
 and the theory for MAE in context of its dependence upon atomic short-range
 compositional order. The penultimate section presents our calculation of the
 dependence of MAE with compositional order in Ni$_{0.8}$Fe$_{0.2}$ and also 
 analyses how much this varies with hypothetical band-filling. We deduce that
 the sensitivity of the high permeabilities of the Ni$_{0.8}$Fe$_{0.2-x}$Cu$_x$ 
 to annealing conditions is itself dependent on the precise composition.
 The final section summarises and concludes.

\section{Magnetic Anisotropy of Metals and Alloys.}

 Magnetic anisotropy of a material derives largely from spin-orbit coupling
 of the electronic structure. Both the 
 origin of this effect, as well as the magnetostatic effects which 
 determine domain structure, can be found from the relativistic 
 generalisation of  spin density functional theory.~\cite{Jansen}
 From the formal starting point of the quantum electrodynamics of 
 electrons interacting with an electromagnetic field, a system's ground 
 state energy is the minimum of a functional of the charge and 
 current densities. This minimisation is achieved, in principle, 
 by the self-consistent solution of a set of Kohn-Sham Dirac equations 
 for independent electrons moving in fields set by the charge and current 
 densities. Once approximations for exchange-correlation and 
 the Gordon decomposition of the current into orbital and spin pieces are 
 made, the spin-orbit coupling effects upon the electronic structure can be 
 represented and the magnetic anisotropy described.  (The magnetostatic 
 shape anisotropy is also described from within the same theoretical 
 framework \cite{Jansen} but not considered for the cubic materials of this paper.) 

 Most theoretical investigations of MAE and calculations of the 
 anisotropy constants $K$ place their emphasis on spin-orbit coupling 
 effects using either perturbation theory or a fully relativistic theory, 
 e.g. \cite{James,Razee+97}. Typically the total energy, or the 
 single-electron contribution to it (if the force theorem is used 
 \cite{arm}), is calculated 
 for two magnetisation directions, ${\bf e}_1$ and ${\bf e}_2$ separately, 
 i.e. $F_{{\bf e}_1}$, $F_{{\bf e}_2}$, and then the MAE, $\Delta F$, is 
 obtained from the difference between them i.e.
\begin{equation}
\Delta F ( {\bf e}_1, {\bf e}_2) = 
\int^{E_{F}^{1}} \varepsilon n_{{\bf e}_{1}}(\varepsilon) d\varepsilon -
\int^{E_{F}^{2}} \varepsilon n_{{\bf e}_{2}}(\varepsilon) d\varepsilon 
\label{eq:MAE}
\end{equation}
 where $E_{F}^{1}$, $E_{F}^{2}$ are the Fermi energies 
 when the system is magnetised along the directions ${\bf e}_{1}$ and 
 ${\bf e}_{2}$ respectively and $n_{{\bf e}_{1(2)}}$ the electronic 
 density of states. However, the MAE is a small part of the total energy 
 of the system, in many cases of the order of $\mu eV$ and it is numerically
 more precise to calculate the difference directly \cite{Razee+97}. 
 We follow this rationale here for the study of soft magnetism and 
 calculate the MAE from
\begin{eqnarray}
\Delta F & =  & - \int^{E_{F_1}} [ N ( \varepsilon;{\bf e}_1 ) -
N (\varepsilon;{\bf e}_2 ) ] d \varepsilon 
- \frac{1}{2} n ( E_{F_2};{\bf e}_2 ) ( E_{F_1} - E_{F_2} )^2 \\ \nonumber
& & + {\mathcal O} ( E_{F_1} - E_{F_2} )^3
\label{eq:de3}
\end{eqnarray}
 where $N (\varepsilon;{\bf e})$ represents the density of states of a
 system magnetised along ${\bf e}$ integrated up to energy $\varepsilon$.
 In most cases, the second term is very small in
 comparison with the first. For a binary component disordered alloy, $A_{1-c}B_c$, 
 the MAE can be written using the Lloyd formula~\cite{Lloyd}
 for the integrated density of states in Eq. (\ref{eq:de3}) to get
\begin{eqnarray}
\Delta F & = & - \frac{1}{ \pi} \Im \int^{E_{F_1}}
d \varepsilon \,
\left[ \frac{1}{ \Omega_{BZ} } \int d {\bf k} \; \ln \| I +
[ t_c^{-1} ({\bf e}_2) - t_c^{-1} ({\bf e}_1) ]
\tau_c ({\bf k};{\bf e}_1) \| \right. \nonumber \\
& + & \left. \frac{}{} (1-c) \{ \ln \| D_A ({\bf e}_1 ) \| -
\ln \| D_A ({\bf e}_2 ) \| \}
+ c \{ \ln \| D_B ({\bf e}_1) \| -
\ln \| D_B ({\bf e}_2) \| \} \frac{}{} \right] \nonumber \\
& - & \frac{1}{2} n ( E_{F_2};{\bf e}_2 )
( E_{F_1} - E_{F_2} )^2 +
{\mathcal O} ( E_{F_1} - E_{F_2} )^3 \label{eq:dealloy}
\end{eqnarray}

 In Eq. (\ref{eq:dealloy}), $ t_c ({\bf e}_1) $ and
 $ t_c ({\bf e}_2) $ are the SPR-KKR-CPA t-matrices
 for magnetisation along ${\bf e}_1$ and ${\bf e}_2$ directions
 respectively and $\tau_c ({\bf k};{\bf e}_1)$ is the 
 scattering path-operator
\begin{equation}
\tau_c ({\bf k};{\bf e}_1) = [ t_c^{-1} ({\bf e}_1 ) -
g ({\bf k}) ]^{-1} \label{eq:tauk}
\end{equation}
 and $\tau_c^{00}$ is its integral over the Brillouin zone. 
 $D_A ({\bf e}_1)$ is found from 
\begin{equation}
D_A ({\bf e}_1) = [ I - \tau_c^{00} ({\bf e}_1) \{ t_c^{-1}
({\bf e}_1)  - t_A^{-1} ({\bf e}_1) \} ]^{-1} \label{eq:da}
\end{equation}
 with similar expressions for $ D_A ({\bf e}_2), \;
 D_B ({\bf e}_1) $, and $ D_B ({\bf e}_2) $. Note that
 $ t_A ({\bf e}_2) $ and $ t_B ({\bf e}_2) $, the single 
 site t-matrices for A and B atoms respectively spin-polarised 
 along ${\bf e}_2$, can be obtained directly from
 $ t_A ({\bf e}_1) $ and $ t_B ({\bf e}_1) $
 respectively by simple rotational transformations.\cite{Razee+97} 
 The numerical accuracy of our SPR-KKR-CPA calculations is
 to within 0.1~$\mu$eV (or 10$^4$~erg/cm$^3$) and thus the scheme 
 is suited for studies of magnetically soft alloys. Full details
 of the method can be found elsewhere.~\cite{Razee+97,Razee+99,BG+EB}

\section{Magnetostriction}

 The magnetostriction constant ${\lambda}_{001}$ represents 
 the relative change of length ($\delta l/l$) measured along
 [001] when an external magnetic field is applied along to the
 direction of observation. For a cubic system ${\lambda}_{001}$
 appears in the expression~\cite{OHandley}
\begin{equation}
 B_1 = - \frac{3}{2} \,
  \lambda_{001} (C_{11} - C_{12}) \, ,
 \label{eq:lambda_001}
\end{equation}
 where, $B_1$ is the rate of change of the MAE, $\Delta F((001), (100))$, with 
 tetragonal strain, $t$, along $[001]$, at $t =0$. 
 $C_{11}$ and $C_{12}$ are the cubic elastic constants which are related 
 to the tetragonal shear ($C^{\prime}$) modulus: 
 $C^{\prime}$ = ($C_{11} - C_{12}$)/2. ${\lambda}_{001}$ of Fe, 
 calculated from this expression using a full-potential DFT method,
 has a large value in comparison with experiment.\cite{Hjortstam,Fast}
 Freeman and co-workers have estimated $C_{11} - C_{12}$ from a fit of their
 calculated total energies to simple quadratic functions of the tetragonal 
 distortion $t$ and calculate  $B_1$ from the gradient 
 of the energy difference  $\Delta F( (0,0,1),(1,0,0))$ with respect to $t$.
 In the case of Fe this approach produces an estimate of ${\lambda}_{001}$ 
 greater than that found experimentally.\cite{Freeman99,WuFreeman}
 Rather than to estimate $C_{11} - C_{12}$ from such total energy 
 calculations, here we assume that it is relatively insensitive to 
 composition and we follow the trends of the magnetostriction 
 from our relativistic electronic structure calculations of $B_1$. 

\section{The magnetic anisotropy and magnetostriction of 
         the prototypical binary Ni$_{0.8}$Fe$_{0.2}$ alloy}

 As the first step, one-electron potentials for the fcc 
 Ni$_{0.8}$Fe$_{0.2}$ and Ni$_{0.75}$Fe$_{0.25}$
 randomly disordered bulk alloys were calculated self-consistently,
 using the scalar relativistic SP-KKR-CPA technique~\cite{Razee+97,Razee+99} 
 and the atomic sphere approximation (ASA).
 The unit-cell volume was fixed at the experimental volume of
 each concentration. Exchange and correlation were accounted
 for using the local spin density approximation (LDA).
 The average spin magnetic moment of Ni$_{1-c}$Fe$_{c}$ alloy decreases 
 slightly with decreasing Fe concentration from 1.14${\mu}_B$ at $c$=0.25 
 to the value of 1.05${\mu}_B$ at $c$=0.2. As $c_{Fe}$ decreases
 between 0.25 $ > c > $ 0.2, the Fe spin magnetic moment
 increases from 2.61${\mu}_B$ to 2.63${\mu}_B$ whereas the Ni spin moment 
 of 0.65${\mu}_B$ is unchanged. These results are in good agreement 
 with both neutron diffraction experiments and previous {\it ab initio} 
 calculations.~\cite{JPS87,SJP87}

 In Fig.~1, we plot the total, spin- and component-resolved 
 densities of states (DOS) of Ni$_{0.8}$Fe$_{0.2}$ together with 
 the band-filling curve $Z_v$. These also concur with previous 
 calculations.~\cite{JPS87,SJP87}
 In common with other late transition metal strong ferromagnets, the 
 majority spin d-states are completely filled and the majority spin DOS 
 associated with the Fe and Ni sites are very similar displaying no 
 evidence of the compositional disorder. This is contrary to the minority
 spin DOS where some of the Fe-related d-states are split away above the
 Ni-related ones.

 Using the fully relativistic SPR-KKR-CPA method~\cite{Razee+97} we 
 calculated the MAE of Ni$_{0.8}$Fe$_{0.2}$ and 
 Ni$_{0.75}$Fe$_{0.25}$ and their dependence  
 on volume conserving tetragonal distortions to obtain $B_1$.
 The results are plotted in the upper panel of Fig.~2 together with
 the experimental MAE value (2.7~$\mu$eV) of bulk fcc Ni. For $c/a =1$, 
 $t_{[0,0,1]} =0$, the easy axes are directed along (111) and  
 the magnitude of the MAE ($\Delta F ((001), (111))$) drops from 1.9~$\mu$eV
 to 0.9~$\mu$eV as $c_{Fe}$ decreases between 0.25 $ < c < $ 0.2. 
 This sharp decrease in the MAE for $c \approx$0.2 is consistent 
 with experiment. 
 The $t$-distortions, which break the cubic symmetry making the 
 [001] and [100] directions non-equivalent, increase the MAE 
 significantly. In Fig.~2, we plot the MAE of Ni$_{1-c}$Fe$_{c}$ 
 as the $c/a$ ratio is altered. The slopes of these graphs around 
 $t=0$ or $c/a =1$ produce the coefficient $B_1$ related to the 
 magnetostriction.
 The two concentrations $c_{Fe}$ were chosen to illustrate the 
 sensitivity of MAE to the Fe content and $t$-deformations.
 $B_1$ is negative for both
 Ni$_{0.8}$Fe$_{0.2}$ and Ni$_{0.75}$Fe$_{0.25}$ and
 $| B_1(c_{Fe}=0.2) |  < | B_1(c_{Fe}=0.25) |$ which is
 consistent with experimental observation that the magnetostriction
 $\lambda$ passes through zero near this composition region.
  
 In the next section we look at the effect of adding non-magnetic
 impurities, Cu and Mo, to nickel-iron alloys in the permalloy 
 composition range. Experimental data suggest that both the MAE 
 and magnetostriction are very sensitive to low concentrations 
 of dopants. We see whether this feature is evident from the 
 electronic structure-based calculations of the MAE and $B_1$.
\section{Magnetic softening in the Ni-Fe-Cu and Ni-Fe-Mo permalloys.}

 Properties of non-magnetic $d$-metals, dissolved substitutionally
 in ferromagnetic hosts, have been well-documented.\cite{Pettifor,Dederichs}  
 If a late transition metal such as Cu is added to permalloy, alongside
 the reduction in the amount of iron, a net decrease in the number of 
 minority spin d-electrons is to be expected. The d-states associated 
 with Cu will hybridise strongly with the nickel-related d-states and
 the Fermi energy for the alloy will be shifted upwards. There is also
 a net reduction in the number of minority spin d-electrons if an early
 transition metal such as Mo is added. The impurities will produce 
 virtual bound states above the ferromagnetic host's majority spin 
 d-bands and the impurity d-levels will hybridise with the host 
 sp-conduction electrons. There will be a further hybridisation with 
 the minority-spin d-electrons in permalloy associated with the Fe sites
 and an anti-parallel moment is expected on the Mo impurity sites. 
 Pushing majority spin states above the Fermi energy by adding Mo thus
 enables more minority spin states to be occupied, $N_{\downarrow}$. 
 If the MAE and magnetostriction vary with $N_{\downarrow}$, as 
 suggested by James {\it et al.}~\cite{James}, Cu and Mo additions to 
 permalloy should have similar effects on these quantities. Our results
 broadly bear this out. The spin magnetic moments, number of valence
 electrons and number of minority spin electrons calculated for various
 ternary Ni-Fe-Cu and Ni-Fe-Mo alloys are given in Table.1.
 
 In Fig.~2 we display  
 $\Delta F((001),(111))$, $\Delta F((001),(110))$ and $\Delta F((001),(111))$ 
 of 5~at.\% Cu(Mo) Ni-Fe-Cu(Mo) alloys, as the $c/a$ ratio is altered.
 The trends in the MAE and $B_1$ for both the Cu- and Mo-doped alloys are 
 roughly similar but there are some differences. 
 In the case of Ni-Fe-Cu, (i) the easy axis 
 is along [111] while the magnitude of MAE never exceeds 1~$\mu$eV
 at $c/a =$1, and (ii) the $B_1$ gradually decreases with  increasing $c_{Fe}$ 
 to a near zero value at $c_{Fe}=$0.225. For the Ni-Fe-Mo system,
 (i) the easy axis along [100] is perpendicular to the 
 magnetisation direction, (ii) the MAE values are nearly zero, and
 (iii) $B_1$ changes the sign when $c_{Fe}$ varies between 0.2 and 0.225.    
 In both cases impurity-doping changes the magnetic anisotropy and
 magnetostriction of Ni-rich Ni-Fe in such a way that enhances their 
 permeability.

 The differences in the total DOS when $c/a$ varies, $\Delta$DOS, 
 of Ni$_{0.75}$Fe$_{0.2}$Cu$_{0.05}$
 and Ni$_{0.75}$Fe$_{0.2}$Mo$_{0.05}$ between the magnetization directions
 along [001] and [111] are plotted in Fig.~3. The comparison with 
 the corresponding $\Delta$DOSs of binary Ni$_{0.8}$Fe$_{0.2}$ illustrates
 the doping effect on the magnetic softening. We note from Fig.~2 that changing
 the composition of permalloy from Ni$_{0.80}$Fe$_{0.20}$ to 
 Ni$_{0.80}$Fe$_{0.15}$Cu$_{0.05}$ or Ni$_{0.80}$Fe$_{0.15}$Mo$_{0.05}$ has 
 switched the sign of the slope, $B_1$, of these lines. This suggests that
 at some intermediate doping level, $B_1$ and therefore the magnetostriction
 should be zero. Before embarking on a course of further calculations for 
 ternary alloys in this composition region we tested a model based upon the 
 parent Ni$_{0.80}$Fe$_{0.20}$ system.
 
 Fig.~4 shows all three magnetic anisotropy energies,
 $\Delta F((001),(111))$,\\ $\Delta F((001),(110))$ and $\Delta F((001),(111))$
 for Ni$_{0.80}$Fe$_{0.20}$ as a function of the tetragonal distortion $t$.
 $B_1$ from all three curves is negative.
 We also see how the MAE and $B_1$ changes when the Fermi energy is shifted
 upwards to mimic roughly the effect of alloying with non-magnetic dopants by
 increasing the number of minority spin-electrons. 
 With the `correct' band-filling (Fermi energy unshifted)
 ($Z_v$=9.6~el.), the sign of $B_1$ is negative as shown also in Fig.~2.
 When $Z_v$=9.85~el. (corresponding to $E_F +$0.14~eV)
 the MAE is tiny and the slope $B_1 \sim 0$ implying a very small
 magnetostriction ($\lambda \simeq 0$). Further increase of $E_F$
 causes $B_1$ to become larger and positive, increasing the magnitude 
 of $\lambda$. We now see if we can reproduce this situation by doping 
 with a non-magnetic transition metal.
 
 Fixing the Ni concentration at $c_{Ni}=$0.8 we calculated
 the MAE and $B_1$ over a narrow range of Cu concentration,  3$< c_{Cu} <$5~at.\%. 
 Fig.~5 shows the results. In line with the scenario described in Fig.~4,
 the MAE and $B_1$ become vanishing small  when between 3 and 4 $\%$ of Cu 
 is added. The ternary alloys Ni$_{0.80}$Fe$_{0.17-\delta}$Cu$_{0.03+\delta}$,
 $0< \delta < 0.01$ are therefore
 very soft magnets, since the easy-axis direction switches 
 and $B_1$, changes sign. 
 Hence, for this narrow composition range of Ni-Fe-Cu, the magnetostriction 
 coefficients, $\lambda \to 0$ while the MAE value remains low enough 
 for this alloy to develop high permeability.

 From the MAE calculations of pure Fe and Ni
 we know that the lattice constant is also an important factor.
 Here we have used the experimental lattice constants and
 therefore, the magnetovolume effect has not been examined.
 Our SPR-KKR-CPA calculations have been used to study  
 the effect of the variation of doping with an early and late transition
 metal on the magnetic anisotropy properties of permalloys and give an 
 explanation of the experimentally observed and well-known facts
 on the preparation of the high permeability materials. In particular we 
 have demonstrated why extremely  
 soft magnetic properties of Ni-Fe-Cu(Mo) are developed for
 rather narrow concentration ranges of the Cu/Mo dopants,
 depending on the Fe content in Ni-rich compositions. Up to now we have
 assumed that the alloys are completely compositionally disordered.
 Preparation procedures for high permeability permalloys, however, do
 instill varying degrees of long- or short-range compositional order which
 alter the magnetic properties significantly. We study this aspect in the
 next sections. 
 
\section{Compositional order and magnetic anisotropy}

 Ordering in alloys can be conveniently and succinctly classified
 in terms of static concentration waves.\cite{Khachaturyan}
 For example,at high temperatures, a binary alloy $A_c B_{1-c}$ where
 the atoms are arranged on a regular array of lattice sites, 
 has each of its sites  occupied by either an $A$- or $B$-type atom 
 with probabilities $c$ and $(1-c)$ respectively. In general, 
 in terms of a set of site-occupation variables $ \{ \xi_i \} $,
 (with $ \xi_i = 1(0) $ when the $i$-th site in the lattice is
 occupied by an $A(B)$-type atom) the thermodynamic average,
 $ \langle \xi_i \rangle $, of the site-occupation variable
 is the concentration $ c_i $ at that site and for the solid
 solution $ c_i = c $ for all sites.
 Below some transition temperature, $T_o$, the system orders or
 phase separates so that a  compositionally modulated alloy forms.
 The temperature-dependent fluctuations of the concentrations
 about the solid solution value $c$,
 $ \{ \delta c_i \} = \{ c_i - c \} $, can be pictured as a
 superposition of static concentration waves,\cite{Khachaturyan,BLG+GMS}
 i.e.,
\[
 c_i = c + \frac12 \sum_{\bf q} \left[
           c_{\bf q} e^{i {\bf q} \cdot {\bf R}_i } +
           c_{\bf q}^\ast e^{-i {\bf q}
           \cdot {\bf R}_i } \right],
\]
 where, $ c_{\bf q} $ are the amplitudes of the concentration
 waves with wave-vectors {\bf q}, and $ {\bf R}_i $ are the
 lattice positions. Usually only a few concentration waves are
 needed to describe a particular ordered structure. For example,
 the CuAu-like $ L1_0 $ tetragonal
 ordered structure is set up by a single concentration wave with
 $ c_{\bf q}=\frac12 $ and $ {\bf q}=(001) $ ({\bf q} is in units
 of $ \frac{2 \pi}{a}$, $ a $ being the lattice parameter). The 
 Cu$_3$Au $L1_2$ ordered phase which Ni$_{0.75}$Fe$_{0.25}$ forms below
 $T=900$K, is set by three waves with $ {\bf q}=(001) $, $(010)$ and
 $(100)$.

 The Free Energy of a partially ordered alloy can be estimated in
 terms of a quantity $S^{(2)}_{ij}$
\[
  S^{(2)}_{jk} ({\bf e}) = - \left.
  \frac{ \partial^2 \Omega (\{ c_i \}; {\bf e} ) }
       { \partial c_j \partial c_k}
   \right|_{ \{ c_i = c \} },
\]
 i.e. a second derivative with respect to concentration of the Grand
 Potential describing the interacting electron system which
 constitutes the alloy.~\cite{BLG+GMS} $S^{(2)}_{ij}$ is formally a direct pair
 correlation function but can loosely be pictured as an effective
 atom-atom interchange energy. It is determined by the electronic
 structure of the disordered phase, the solid solution. The atomic
 short-range order, $ \alpha_{ij} = \beta [ \langle \xi_i \xi_j \rangle -
\langle \xi_i \rangle \langle \xi_j \rangle]$, whose lattice
 Fourier transform $\alpha ({\bf q},T)$ can be measured
 by diffuse scattering experiments, is related directly via
 $\alpha ({\bf q},T) = \beta c (1-c)/( 1 - \beta c (1-c) S^{(2)} ({\bf q}))$
 where $S^{(2)} ({\bf q})$ is the lattice Fourier transform of
 $S^{(2)}_{ij}$. The spinodal transition temperature $ T_o $, below
 which the alloy orders into a structure characterized by the
 concentration wave-vector ${\bf q}_{max}$, is determined by
 $ S^{(2)} ({\bf q}_{max}) $, where $ {\bf q}_{max}$ is
 the value at which $S^{(2)} ({\bf q})$ is maximal
 ($ {\bf q}_{max}=(0,0,1)$ for $ L1_0 $ order).
 We can write,\cite{BLG+GMS,JBS+DDJ+FJP}
 $T_o = c(1-c) S^{(2)} ({\bf q}_{max})/k_B$. Calculations of
 $S^{(2)} ({\bf q})$ then can provide a quantitative description
 of the propensity of an alloy to order when thermally annealed.

 The MAE of a ferromagnetic alloy with a compositional modulation
 can be defined in terms of the difference between the free energies
 of the system magnetised along two different directions
 ${\bf e}_1$ and ${\bf e}_2$. If the modulation is specified
 by a concentration wave of amplitude $ c_{\bf q} $ and wavevector
 ${\bf q}$, this difference can be approximately written as
\begin{equation}
  \Delta F ({\bf e}_1, {\bf e}_2)) \approx
 \Delta F^{dis.}({\bf e}_1, {\bf e}_2)
        - K ({\bf q};{\bf e}_1, {\bf e}_2) | c_{\bf q} |^2
 \label{eq:kq}
\end{equation}
 where $K ({\bf q};{\bf e}_1, {\bf e}_2) = \frac{1}{2}[ S^{(2)}
 ({\bf q}; {\bf e}_1 ) - S^{(2)} ({\bf q}; {\bf e}_2 )]$, half
 the difference between the direct correlation function
 $S^{(2)} ({\bf q})$ for the ferromagnetic alloy magnetised along
 ${\bf e}_1$ and along ${\bf e}_2$. $\Delta F^{dis.}$ is
 the MAE of the completely disordered alloy~\cite{Razee+97} which we 
 have calculated earlier in this paper for Ni-Fe, Ni-Fe-Cu and Ni-Fe-Mo.
 For an alloy with atomic short range order $\alpha ({\bf q}; T)$
 the MAE is expressed
\begin{equation}
 \Delta F({\bf e}_1, {\bf e}_2) \approx
 \Delta F^{dis.}({\bf e}_1, {\bf e}_2)
    -\frac{1}{V_{BZ}} k_B T \int d {\bf q}'
  K ({\bf q}'; {\bf e}_1, {\bf e}_2) \alpha ({\bf q}',T; {\bf e}_1)
\end{equation}
 Now $\alpha ({\bf q},T; {\bf e}_1)$ is a structured function of
 ${\bf q}$ with peaks located at ${\bf q}_{max}$, wavevectors of
 the concentration waves which characterise the ordered phase
 the alloy can form at low temperatures at equilibrium.
 Just above $T_o$ where the ASRO is pronounced the second term
 becomes $ \approx -c(1-c) K({\bf q}_{max}; {\bf e}_1, {\bf e}_2)$.
 Full details on how the dependence of MAE on compositional ordering
 can be calculated from ab-initio electronic structure calculations
 can be found in references \cite{Razee+97,Razee+99}.
 Another aspect of this approach can be obtained from
 calculating $ S^{(2)} ({\bf q}; {\bf e} ) $ for different
 {\bf q}-vectors, whilst keeping the magnetic field and
 magnetization direction fixed and gives a quantitative description
 of the phenomenon of magnetic annealing \cite{Razee+99} in
 which an applied magnetic field can cause directional
 compositional ordering. For a magnetically soft alloy such as
 the 75~\% Ni permalloy this leads to a significant uniaxial
 anisotropy.\cite{Razee+97} In addition to permalloy, we have also used
 this approach to investigate the relation of MAE to compositional
 order in Fe$_{0.5}$Co$_{0.5}$, Co$_{0.5}$Pt$_{0.5}$
 and Fe$_{0.5}$Pd$_{0.5}$.\cite{Razee+2001}

 In the next section we use this approach to investigate how the effect
 of compositional order on the magnetic anisotropy of permalloy
 Ni$_{0.80}$Fe$_{0.20}$ varies as the number of minority spin electrons
 is increased. Once again this can be taken as a simple model of the 
 influence of adding non-magnetic impurities.

\section{The dependence of the magnetocrystalline anisotropy of
         Ni$_{0.8}$Fe$_{0.2}$ upon chemical order}

 As expected from its inherent cubic symmetry and confirmed by our 
 calculations, completely disordered fcc-Ni$_{0.8}$Fe$_{0.2}$ has
 a tiny magnetic anisotropy ($ \Delta F^{dis.} ((0,0,1), (1,1,1))
 < 1 \mu$eV). This is demonstrated in the upper panel of Fig.~6 and
 discussed earlier. The magnetic anisotropy is further reduced by
 shifting the Fermi energy upwards so that $N_{\downarrow}$ 
 is increased. The lower panel of Fig.~6 shows what happens to 
 $K ({\bf q}'; (001),(100))$ and $K ({\bf q}'; (001),(111))$ for similar
 circumstances. For the `correct' $E_F$, (Z$_{\nu}$ =9.6 electrons), the
 values suggest a significant enhancement of the magnetic anisotropy if
 there is directional  long- or short-range order in Ni$_{0.8}$Fe$_{0.2}$.
 As $E_F$ is shifted upwards so that Z$_{\nu}$ =9.7, 
 $K ({\bf q}'; (001),(100))$ and $K ({\bf q}'; (001),(111))$ double in size
 but then drop sharply to zero when the band-filling reaches 9.83.  The 
 MAE and $B_1$ of the disordered alloy are also very small at this 
 band-filling. We can deduce therefore that a ternary alloy with composition
 near Ni$_{0.80}$Fe$_{0.17}$Cu$_{0.03}$, compatible with this band-filling,
 may have high permeability which is also rather insensitive to its preparation
 conditions. Fig.~6, moreover, shows that this insensitivity is expected 
 only for a very narrow composition range.

\section{Summary}

 Calculations of the magnetic anisotropy and the trends for 
 magnetostriction in fcc 80~\% Ni based Ni-Fe permalloys have been 
 carried out, using the fully relativistic spin-polarized KKR method 
 within the CPA. 
 The reliability of this method to model magnetic anisotropy
 in soft magnetic materials has been demonstrated. The trends found 
 in the MAE calculations 
 for Ni$_{0.8}$Fe$_{0.2}$ are in a good agreement with those found in 
 experiments. 
 A description of magnetic softening caused by doping with non-magnetic metals
 Cu and Mo has been given. We have linked these results to further
 calculations for Ni$_{0.8}$Fe$_{0.2}$ where the number of minority-spin
 electrons has been artificially increased. Our estimates of the effects
 of atomic short- and long-range chemical order on the magnetic 
 anisotropy of these permalloys show strong variation with composition.
 Pooling the results, we have managed to follow the 
 well-known empirical trends and to 
 show how  a potentially perfect permalloy might be found with zero magnetic 
 anisotropy and zero magnetostriction and which is also rather
 insensitive to its preparation conditions.

\section{Acknowledgments}

 We acknowledge support from the EPSRC (UK) and computing resources
 at the Centre for Scientific Computing, University of Warwick.


\begin{figure}
\begin{center}
\begin{minipage}{8.0in}
\epsfxsize=5.0in
\epsfysize=5.0in
{\epsfbox{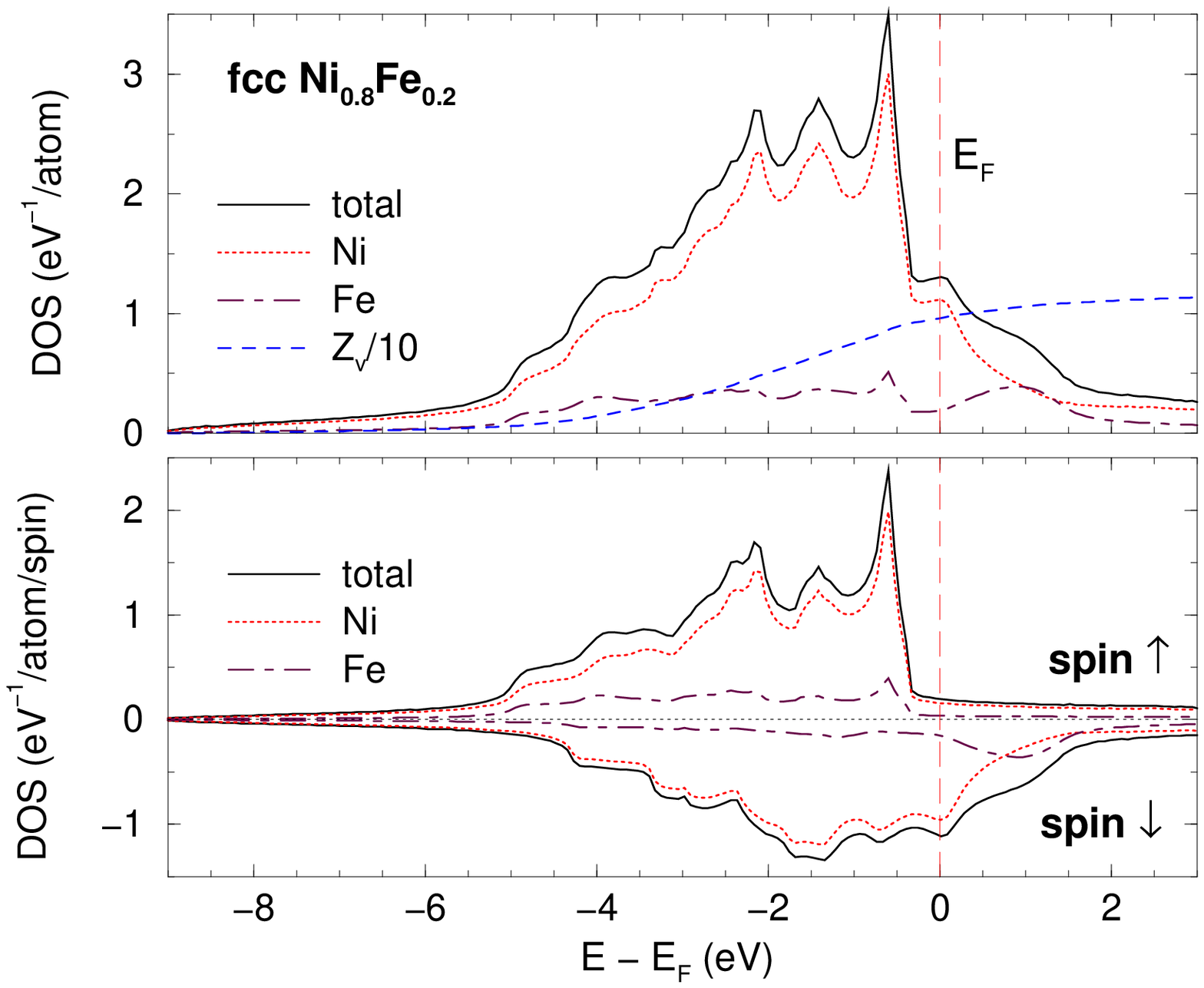}}
\end{minipage}
\end{center}

\vskip 2cm

\caption{The component-resolved and total DOS of  
 Ni$_{0.8}$Fe$_{0.2}$ are shown in the upper 
 panel (a) together with the band-filling curve $Z_v$. 
 In the (b) panel, the spin-projected DOS are shown.}
\label{DOS_Ni80Fe20}
\end{figure}
\begin{figure}
\begin{center}
\begin{minipage}{8.0in}
\epsfxsize=5.0in
\epsfysize=5.0in
{\epsfbox{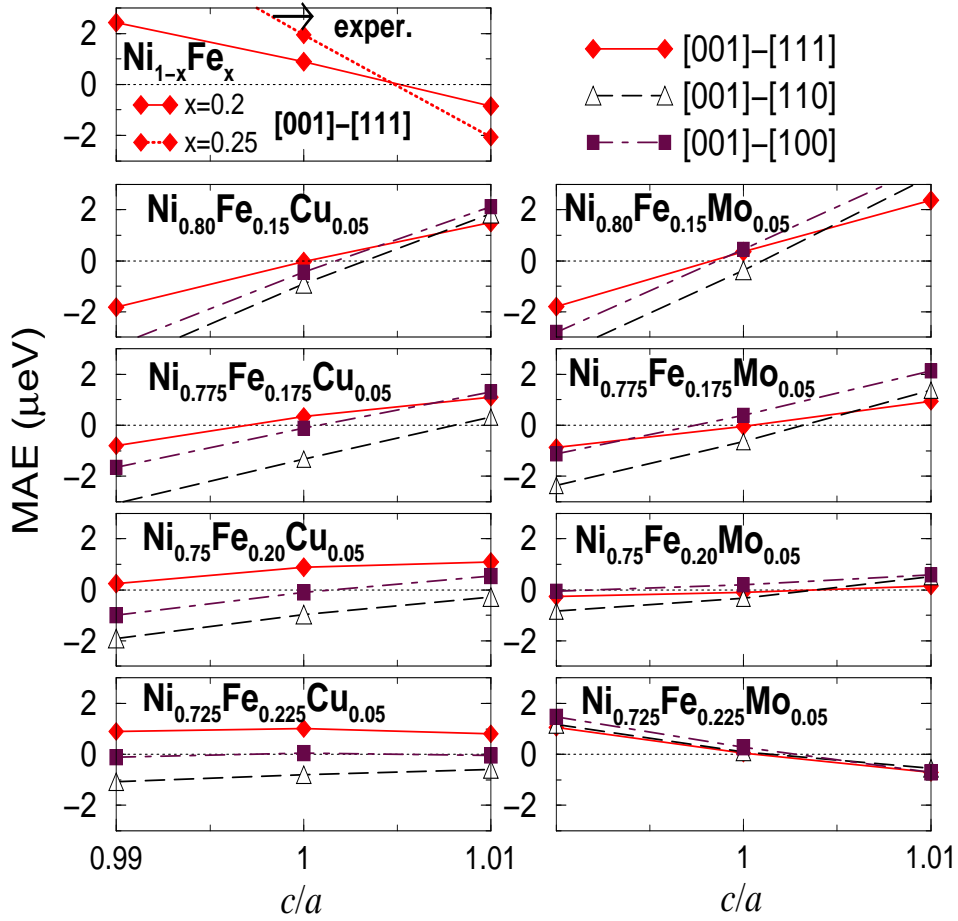}}
\end{minipage}
\end{center}

\vskip 2cm

\caption{The MAE, $\Delta F((001),{\bf e}_2)$, with ${\bf e}_2 =
(100)$, $(110)$ and $(111)$ of randomly
 disordered fct Ni$_{1-c}$Fe$_c$, calculated by SPR-KKR-CPA,
 is shown as a function of tetragonal $c/a$ variations in the 
 upper (left) panel. The experimental MAE value of bulk fcc Ni
 is marked by an arrow.
 In the following four left (right) panels, we plot the 
 variations of $\Delta F$'s with respect to $c/a$ for the ternary
 Ni-Fe-Cu(Mo) alloys doped with 5~at.\% Cu (Mo).  
 }
\label{MCA_FULL}
\end{figure}
\begin{figure} 
\begin{center} 
\begin{minipage}{8.0in} 
\epsfxsize=5.0in 
\epsfysize=5.0in 
{\epsfbox{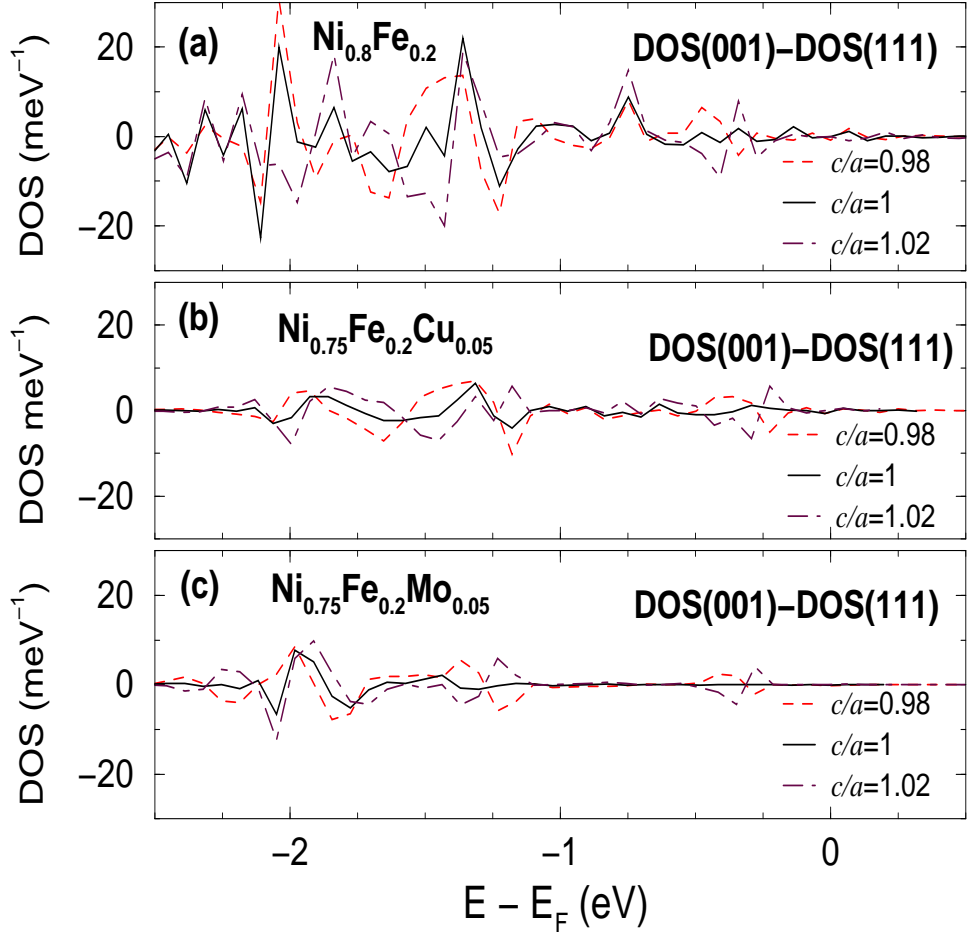}} 
\end{minipage}
\end{center}

\vskip 2cm
\caption{The differences in the total DOS of Ni$_{0.8}$Fe$_{0.2}$
 between the magnetization directions along [001] and [111]
 are plotted in panel (a) for perfect cubic structure, 
 the tetragonal compression ($c/a =$0.98) and elongation 
 ($c/a =$0.98). The corresponding
 $\Delta$DOS of Ni$_{0.75}$Fe$_{0.2}$Cu$_{0.05}$
 and Ni$_{0.75}$Fe$_{0.2}$Mo$_{0.05}$ are shown in the (b) 
 and (c) panels, respectively.  
}
\label{DOS-DIFF} 
\end{figure} 
\begin{figure} 
\begin{center}
\begin{minipage}{8.0in}
\epsfxsize=5.0in 
\epsfysize=5.0in
{\epsfbox{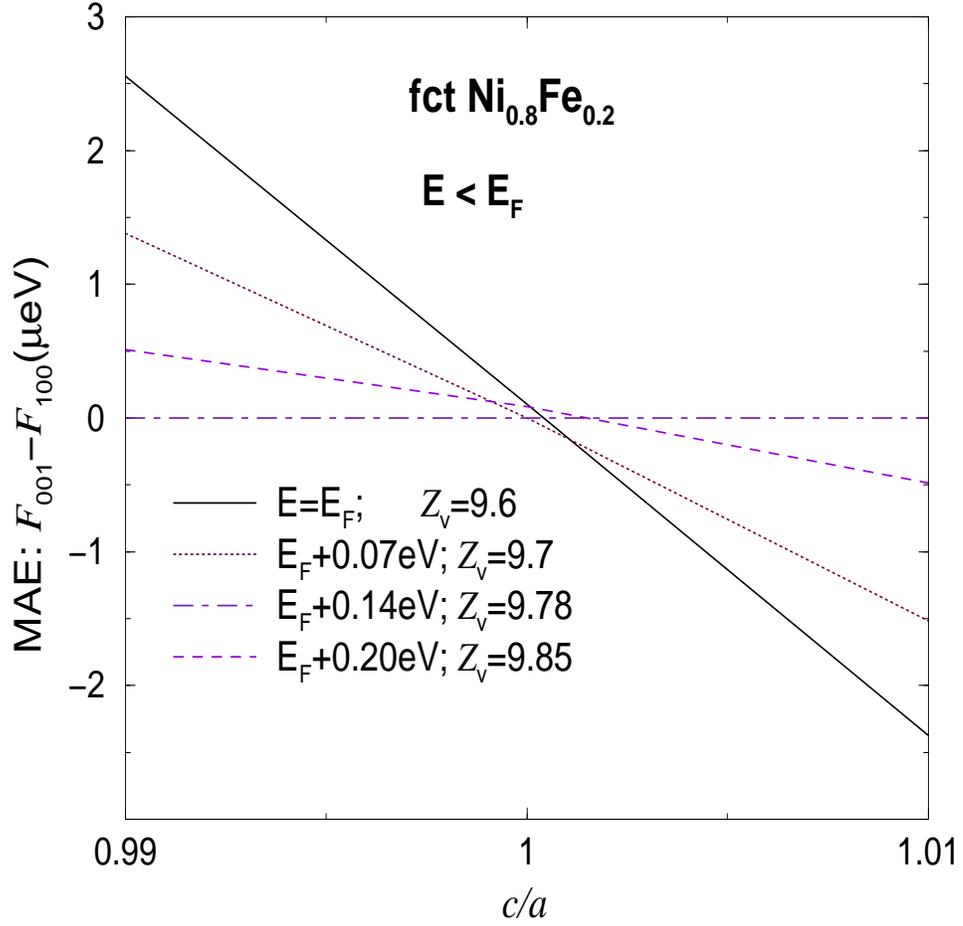}}
\end{minipage} 
\end{center}

\vskip 2cm 
\caption{The MAE, $\Delta F((001), (100))$, of 
 Ni$_{0.8}$Fe$_{0.2}$ as a function
 of the $t$-deformations for several different band-fillings.
 } 
\label{EF_CHANGE}
\end{figure} 
\begin{figure}
\begin{center}
\begin{minipage}{8.0in}
\epsfxsize=5.0in
\epsfysize=5.0in
{\epsfbox{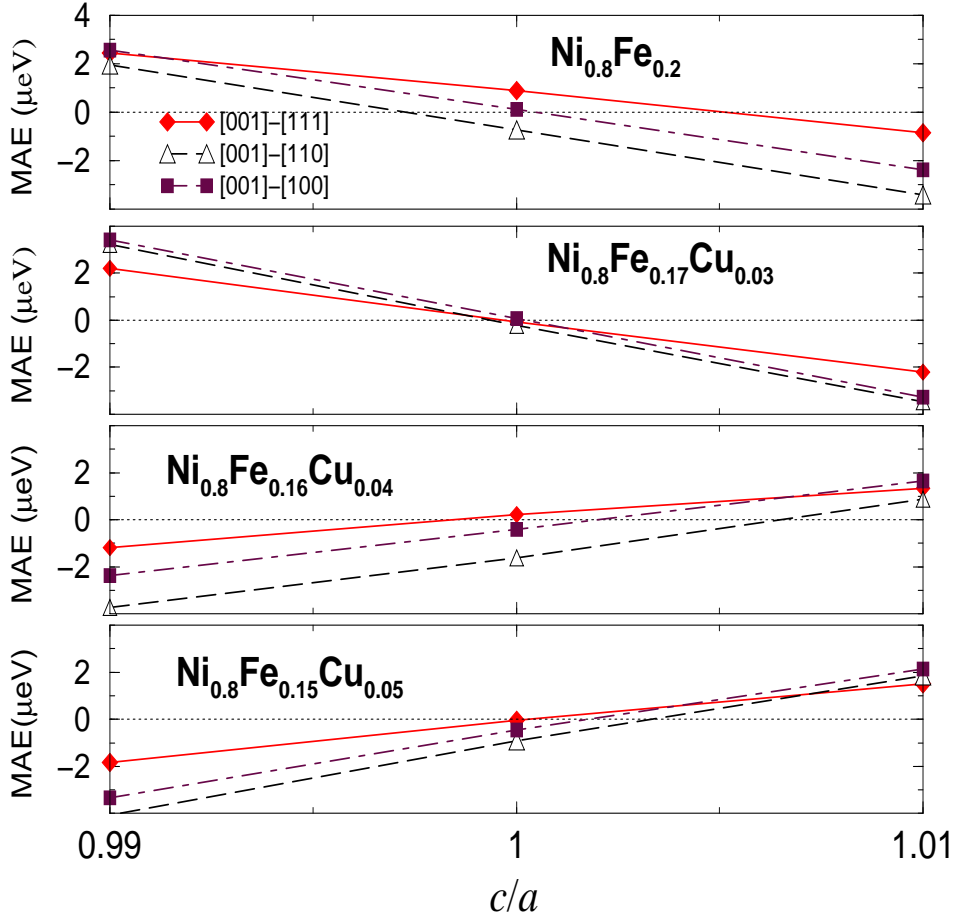}}
\end{minipage}
\end{center}

\vskip 2cm

\caption{The MAE $\Delta F((001),{\bf e}_2)$ for ${\bf e}_2= (111)$,
 $(110)$ and $(100)$ as a function of
 tetragonal deformations for the 80~\% Ni based permalloys.
 The $\Delta F$'s of the Ni$_{0.8}$Fe$_{0.2}$ binary alloy
 are plotted in the upper panel. In the three lower panels,
 we show how the sign of the slope, $B_1$, is changed
 when concentration of dopant (Cu) varies between 3 and 5~at.\%.   
 }
\label{MCA_v_Cu}
\end{figure}
\begin{figure} 
\begin{center}
\begin{minipage}{8.0in} 
\epsfxsize=5.0in
\epsfysize=5.0in 
{\epsfbox{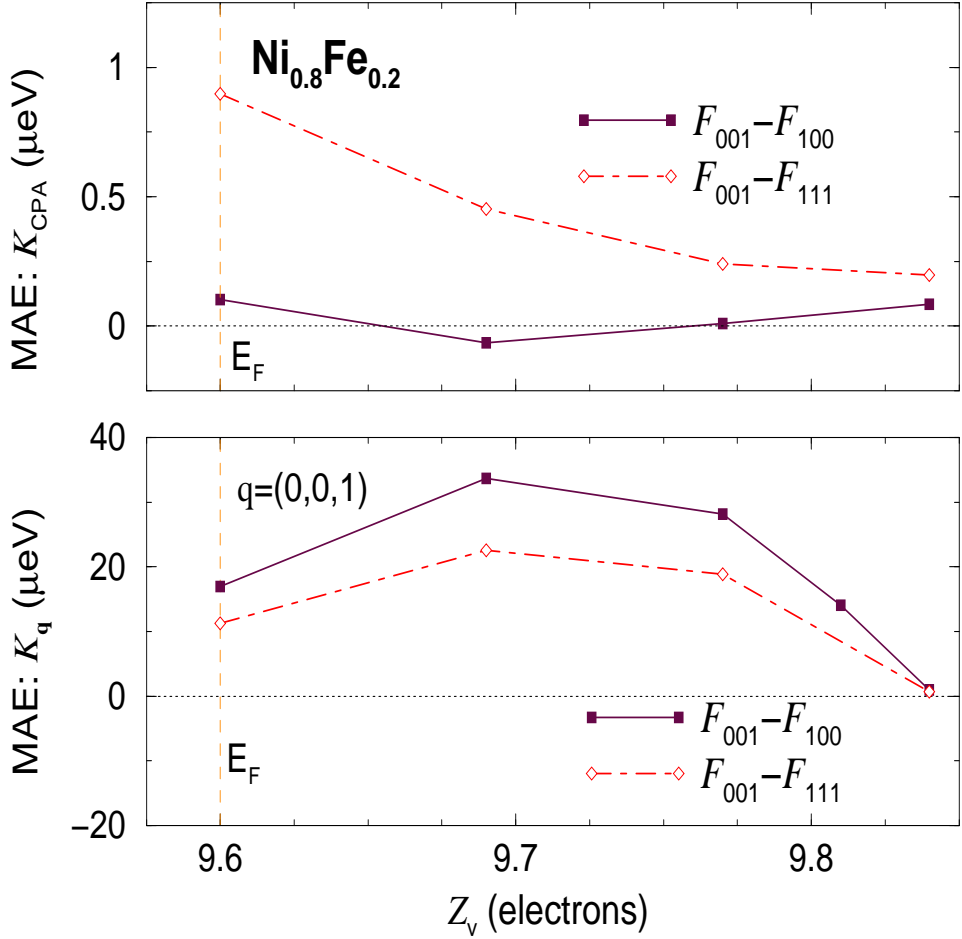}} 
\end{minipage} 
\end{center}

\vskip 2cm 
\caption{The $\Delta F((001), {\bf e}_2)$'s for ${\bf e}_2 = (100)$ and
$(111)$ of the randomly 
 disordered Ni$_{0.8}$Fe$_{0.2}$ alloy are shown in the upper 
 panel as a function of the band-filling parameter $Z_v$.
 In the lower panel, the corresponding variation of the $\Delta F$'s 
 with composition modulation, i.e. $K({\bf q}; (001), {\bf e}_2)$  
 calculated for the modulation ${\bf q} = (0,0,1) $ are plotted as functions 
 of $Z_v$. 
}
\label{K2_EF_CHANGE} 
\end{figure}
\begin{table}
\begin{center}
\begin{minipage}{8.0in}
\begin{tabular}{|c|c|c|c|c|c|c|}
\hline
 & ${\mu}^{Ni}$ & ${\mu}^{Fe}$   & ${\mu}^{Cu(Mo)}$  & $\bar{\mu}$ & $Z_{\nu}$ &
$N_{\downarrow}$ \\
\hline
 Ni$_{0.725}$Fe$_{0.225}$Cu$_{0.05}$ & 0.58 & 2.59 &  0.06 & 1.01 & 9.60 & 8.59 \\
 Ni$_{0.725}$Fe$_{0.225}$Mo$_{0.05}$ & 0.39 & 2.30 & -0.12 & 0.79 & 9.35 & 8.56 \\
\hline
 Ni$_{0.75}$Fe$_{0.25}$           & 0.65 & 2.61 &   --   & 1.14 & 9.50 & 8.36 \\
 Ni$_{0.75}$Fe$_{0.2}$Cu$_{0.05}$ & 0.57 & 2.63 &  0.05  & 0.96 & 9.65 & 8.69 \\
 Ni$_{0.75}$Fe$_{0.2}$Mo$_{0.05}$ & 0.42 & 2.37 & -0.08  & 0.79 & 9.40 & 8.61 \\
\hline
 Ni$_{0.775}$Fe$_{0.175}$Cu$_{0.05}$ & 0.58 & 2.62 &  0.05 & 0.91 & 9.70 & 8.79 \\
 Ni$_{0.775}$Fe$_{0.175}$Mo$_{0.05}$ & 0.41 & 2.39 & -0.08 & 0.73 & 9.45 & 8.72 \\
\hline
 Ni$_{0.8}$Fe$_{0.2}$             & 0.65 & 2.63 &   --   & 1.05  & 9.60 & 8.55 \\
 Ni$_{0.8}$Fe$_{0.15}$Cu$_{0.05}$ & 0.58 & 2.66 &  0.04  & 0.87  & 9.75 & 8.88 \\
 Ni$_{0.8}$Fe$_{0.15}$Mo$_{0.05}$ & 0.40 & 2.40 & -0.07  & 0.68  & 9.50 & 8.82 \\
\hline
\end{tabular}
\end{minipage}
\end{center}
\vskip 2cm
\caption{The component-resolved $\mu^{\alpha}$, average spin magnetic
         moments $\bar{\mu}$ (in ${\mu}_B$), band-filling $Z_{\nu}$ and
         minority spin electrons $N_{\downarrow} = Z_{\nu}- \bar{\mu}$
         in 80~\% based Ni permalloys. }
\label{table1}
\end{table}


\begin{thebibliography}{99}

\bibitem{Tremolet} E. du Tremol\'et de Lacheisserie,
 {\it Magnetostriction: Theory and Applications of
 Magnetoelasticity} (CRC Press, Boca Raton, 1993).

\bibitem{Chikazumi} S.Chikazumi, {\it Physics of Ferromagnetism},
(Oxford Science Publications, 1999).

\bibitem{Wakiyama} T.Wakiyama, {\it Physics and Engineering Applications of Magnetism},
 ed. Y.Ishikawa and N.Miura, Springer Series in Solid State Sciences {\bf 92}, 133, 
 (1991).

\bibitem{Pfeifer} F.Pfeifer and C.Radeloff,
 J. Magn. Magn. Mater. {\bf 19}, 190 (1980). 

\bibitem{Major} H.H.Scholefield, R.V.Major, B.Gibson and A.P.Martin, 
 Brit.J.Appl.Phys. {\bf 18}, 41, (1967).

\bibitem{OSRD-PRB-2004} S.Ostanin, J.B.Staunton, S.S.A.Razee, 
 C.~Demangeat, B.~Ginatempo and Ezio Bruno,
 Phys. Rev. B {\bf 69}, 064425 (2004).

\bibitem{Razee+97} S.S.A.Razee, J.B.Staunton, and F.J.Pinski,
 Phys. Rev. B {\bf 56}, 8082 (1997).

\bibitem{Razee+99} S.S.A.Razee, J.B.Staunton, B.Ginatempo, F.J.Pinski,
 and E.~Bruno, Phys. Rev. Lett. {\bf 82}, 5369 (1999).

\bibitem{James} P.James, O.Eriksson, O.Hjortstam, B.Johansson, 
 L.Nordstrom, Appl. Phys. Lett. {\bf 76}, 915, (2000).

\bibitem{Jansen} H.J.F.Jansen, Phys. Rev. B {\bf 59}, 4699 (1999).

\bibitem{arm} A.R.Mackintosh and O.K.Andersen, in
 {\it Electrons at the Fermi Surface}, edited by M.Springfold
  (Cambridge University Press, Cambridge, 1980); M.Weinert,
   R.E.Watson, and J.W.Davenport,
   Phys. Rev. B {\bf 32}, 2115 (1985).

\bibitem{Lloyd} P.Lloyd and P.V.Smith, Adv.Phys. {\bf 21}, 69, (1972).

\bibitem{BG+EB} B.Ginatempo and E.Bruno, Phys.Rev.B {\bf 55}, 12946, (1997).

\bibitem{OHandley} R.C.O'Handley, {\it Modern Magnetic Materials Principles
 and Applications} (Wiley, 2000).

\bibitem{Hjortstam} O.Hjortstam, K.Baberschke, J.M.Wills, B.Johansson, 
 and O.Eriksson, Phys. Rev. B {\bf 55},  15026 (1997).

\bibitem{Fast} L.Fast, J.M.Wills, B.Johansson, and O.Eriksson,
 Phys. Rev. B  {\bf 51}, 17431 (1995).

\bibitem{Freeman99} A.J.Freeman, R.Q.Wu, M.Y.Kim, and V.I.Gavrilenko,
  J. Magn. Magn. Mater. {\bf 203}, 1 (1999).

\bibitem{WuFreeman} R.Wu and A.J.Freeman, J. Appl. Phys.
 {\bf 79}, 6209 (1996).

\bibitem{JPS87}
 D.D.Johnson, F.J.Pinski, and J.B.Staunton,
 J. Appl. Phys. {\bf 61}, 3715 (1987).

\bibitem{SJP87} J.B.Staunton, D.D.Johnson and B.L.Gyorffy,
  J. Appl. Phys. {\bf 61}, 3693 (1987). 

\bibitem{Pettifor} D.G.Pettifor, {\it Bonding and Structure of
 Molecules and Solids} (Oxford University Press, 1995).

\bibitem{Dederichs} P.H.Dederichs {\it et al.}, J.Mag.Magn.Mat. {\bf 100},
1-3,241-260, (1991); R.Zeller, J.Phys.F {\bf 17}, 2123, (1987).

\bibitem{Khachaturyan} A.G.Khachaturyan,
 {\it Theory of Structural Transformations in Solids}
 (Wiley, N.Y., 1983), p.39.

\bibitem{BLG+GMS} B.L.Gyorffy and G.M.Stocks,
 Phys. Rev. Lett. {\bf 50}, 374 (1983).

\bibitem{JBS+DDJ+FJP} J.B.Staunton, D.D.Johnson and F.J.Pinski,
 Phys. Rev. B  {\bf 50}, 1450 (1994).

\bibitem{Razee+2001} S.S.A.Razee, J.B. Staunton, B. Ginatempo, E. Bruno,
and F.J. Pinski,  Phys. Rev. B {\bf 64}, 014411 (2001).

\end{thebibliography}
\end{document}